\documentclass[11pt]{article}
\usepackage{amsmath}
\usepackage{amsfonts}
\usepackage{amssymb}
\usepackage{epsfig}
\usepackage{times}
\newcommand{\ms}{\Delta m^2_{\odot}}
\newcommand{\ma}{\Delta m^2_{atm}}
\newcommand{\ts}{\sin^2 2\theta_{\odot}}
\newcommand{\beq}{\begin{equation}}
\newcommand{\eeq}{\end{equation}}
\newcommand{\bea}{\begin{eqnarray}}
\newcommand{\eea}{\end{eqnarray}}
\def\ltap{\ \raisebox{-.4ex}{\rlap{$\sim$}} \raisebox{.4ex}{$<$}\ }
\def\gtap{\ \raisebox{-.4ex}{\rlap{$\sim$}} \raisebox{.4ex}{$>$}\ }

\makeatletter
\renewcommand{\section}{\@startsection
{section}%
{1}%
{0em}%
{-1mm}%
{1mm}%
{\Large\bfseries}}%
\renewcommand{\subsection}{\@startsection
{subsection}%
{2}%
{0em}%
{-1mm}%
{1mm}%
{\bfseries}}%
\begin{document}
%
\begin{titlepage}
November 2001 \hfill 
\hfill  Ref.~SISSA 88/2001/EP
\begin{flushright} hep-ph/0112074 \end{flushright}
\vskip 3.5cm
{\baselineskip 17pt
\begin{center}
{\bf THE LMA MSW SOLUTION OF THE SOLAR NEUTRINO PROBLEM,
\break INVERTED NEUTRINO MASS HIERARCHY AND
REACTOR NEUTRINO EXPERIMENTS}
\end{center}
}
\vskip .3cm
\centerline{
 \ S.T. Petcov
     \footnote[1]{SISSA/INFN, Via Beirut 2-4, I-34014 Trieste, Italy.}
$\raisebox{-.2ex}{}$$^,$
     \footnote[2]{Also at: Institute of Nuclear Research and
Nuclear Energy, Bulgarian Academy of Sciences, 1784 Sofia, Bulgaria.}
and\ M. Piai~$^{1}$
     }
\vskip 1.5cm
%
%
{\bf Abstract.} In the context of 
three-neutrino oscillations,
we study the possibility of using antineutrinos
from nuclear reactors to explore the 
$10^{-4}~{\rm eV^2} < \ms \ltap 8\times 10^{-4}~{\rm eV^2}$
region of the LMA MSW solution of 
the solar neutrino problem and measure
$\ms$ with high precision.
The KamLAND experiment is 
not expected to  
determine $\ms$
if the latter happens to lie 
in the indicated region.
By analysing both 
the total event rate suppression and 
the energy spectrum distortion
caused by $\bar{\nu}_e$ oscillations in vacuum,
we show that the optimal baseline of
such an experiment 
is $L \sim (20 - 25)$ km. 
Furthermore, for
$10^{-4}~{\rm eV^2} < \ms \ltap 5\times 10^{-4}~{\rm eV^2}$,
the same experiment might be used to try to 
distinguish between the two possible 
types of neutrino mass spectrum - 
with normal or with inverted hierarchy, 
by exploring the effect of interference 
between the atmospheric- and 
solar- $\Delta m^2$  
driven oscillations;  
for larger values of $\ms$ 
not exceeding $8.0\times 10^{-4}$ eV$^2$,
a shorter baseline, $L \cong 10$ km, would be
needed for the purpose. 
The indicated interference effect
modifies in a characteristic way 
the energy spectrum of detected events.
Distinguishing between
the two types of neutrino mass spectrum
requires, however, 
a high precision determination
of the atmospheric $\Delta m^2$, a 
sufficiently large $\sin^2\theta$ and
a non-maximal $\sin^22\theta_{\odot}$,
where $\theta$ and $\theta_{\odot}$
are the mixing angles respectively 
limited by the CHOOZ and Palo Verde data and
characterizing the solar neutrino
oscillations. It also requires
a relatively high precision measurement of the
positron spectrum in the reaction
$\bar{\nu}_e + p \rightarrow e^{+} + n$.


\smallskip

\vfill
\vskip 2.5cm
\centerline{
\epsfig{file=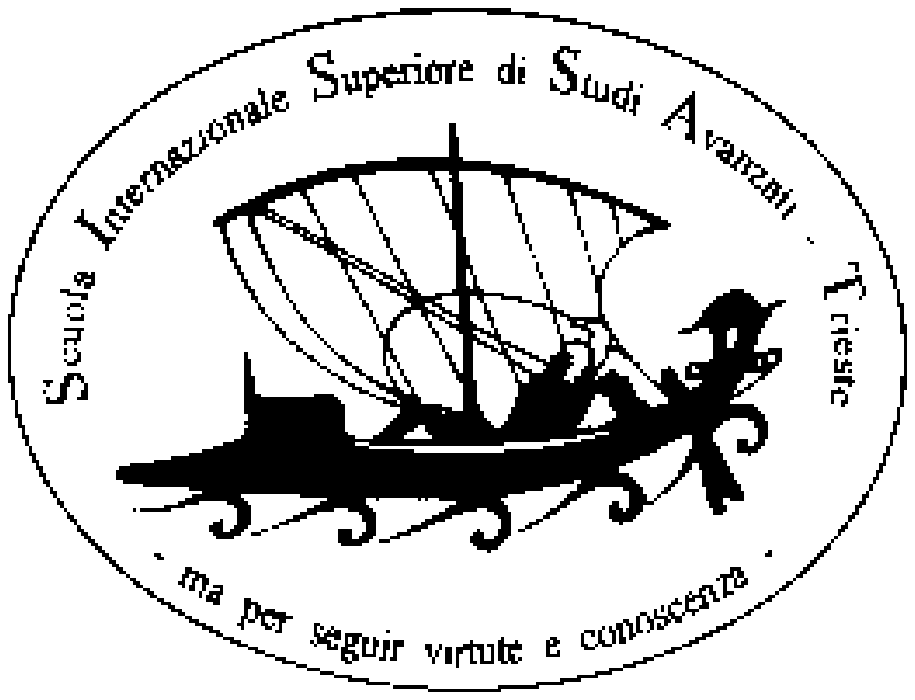,height=2.3truecm,width=3.5truecm}
}
\end{titlepage}
%
%
\section{Introduction}
\label{section:introduction}
\vspace{0.1cm}

\indent \hskip 1.0truecm In recent years the experiments 
with solar and atmospheric neutrinos
collected strong evidences in favor of the existence of 
oscillations between the  
flavour neutrinos, 
$\nu_e$, $\nu_{\mu}$ and $\nu_{\tau}$.  
Further progress in our 
understanding of the neutrino 
mixing and oscillations requires, in particular, 
precise measurements of the parameters 
entering into the oscillation probabilities - 
the neutrino mass-squared differences
and mixing angles, and the reconstruction 
of the neutrino mass spectrum.

\indent  \hskip 1.0truecm The atmospheric neutrino data can be explained by
dominant $\nu_{\mu} \rightarrow \nu_{\tau}$ and 
$\bar{\nu}_{\mu} \rightarrow \bar{\nu}_{\tau}$ 
oscillations, characterized by large, possibly maximal,
mixing, and a mass squared difference, $\ma$,
having a value 
in the range \cite{SKatm} (99\% C.L.):
\bea
 1.3 \times 10^{-3} {\mbox{eV}}^2 \ltap |\ma| \ltap
 5 \times 10^{-3} {\mbox{eV}}^2.
\label{atmlimit}
\eea
%
\indent The first results from the Sudbury Neutrino Observatory 
(SNO)~\cite{SNO}, combined with
the mean event rate data from the 
Super-Kamiokande (SK) experiment~\cite{SKsol},
provide a very strong evidence for
oscillations of the solar neutrinos 
\cite{BargerSNO} - \cite{KSSNO}.
Global analyses of the solar neutrino data,
including the SNO results and the SK data
on the $e^{-}-$spectrum and day-night asymmetry, 
show that the data favor the large mixing 
angle (LMA) MSW solution
of the solar neutrino problem,
with the corresponding neutrino 
mixing parameter $\ts$ and
mass-squared difference $\ms$ 
lying in the regions (99.73\% C.L.):
\bea
 2 \times 10^{-5} {\mbox{eV}}^2 \ltap \ms 
\ltap 8 \times 10^{-4} {\mbox{eV}}^2
\label{sollimit}
\\
0.6 \leq \ts \leq 1.
\label{thetalimit}
\eea

The best fit value of $\ms$ found in the 
independent analyses 
\cite{FogliSNO,ConchaSNO,GoswaSNO,GiuntiSNO}
is spread in the interval 
$(4.3 - 6.3)\times 10^{-5}~{\rm eV^2}$.
The results obtained 
in \cite{FogliSNO,ConchaSNO,GoswaSNO,GiuntiSNO}
show that values of $\ms > 10^{-4}~{\rm eV^2}$
are allowed already at 90\% C.L. 
Values of $\cos 2\theta_{\odot} < 0$ 
(for $\ms >0$) are disfavored 
by the data. 

\indent  \hskip 1.0truecm Important constraints on 
the oscillations of electron (anti-)neutrinos,
which play a significant role
in our current understanding
of the possible patterns of oscillations of the 
three flavour neutrinos and 
anti-neutrinos,
were obtained in the CHOOZ and Palo Verde 
disappearance experiments 
with reactor $\bar{\nu_e}$ \cite{CHOOZ,PaloV}.
The CHOOZ and Palo Verde experiments
were sensitive to values of neutrino mass 
squared difference
$\Delta m^2 \gtrsim 10^{-3}~{\rm eV^2}$,
which includes the corresponding 
atmospheric neutrino region, eq. (1).
No disappearance of the reactor  $\bar{\nu}_e$
was observed. Performing a
two-neutrino oscillation
analysis, the following rather stringent
upper bound on the value of the
corresponding mixing angle, $\theta$,
was obtained by the CHOOZ collaboration
\footnote{The  possibility of large
$\sin^2  \theta > 0.9$
which is admitted by the CHOOZ data alone
is incompatible with the neutrino oscillation
interpretation of the solar neutrino deficit
(see, e.g., \cite{BBGK9596,BGG99})}
\cite{CHOOZ} at 95\% C.L. for
$\Delta m^2 \geq 1.5 \times 10^{-3} \mathrm{eV}^2$:
\begin{equation}
\sin^{2} \theta < 0.09.
\label{chooz1}
\end{equation}
%
\noindent The precise upper limit in eq. (\ref{chooz1})
is $\Delta m^2$-dependent:
it is a decreasing function of $\Delta m^2$ as
$\Delta m^2$ increases up to
$\Delta m^2 \simeq 6 \cdot 10^{-3}~\mathrm{eV}^2$
with a minimum value
$\sin^{2} \theta \simeq 10^{-2}$.
The upper limit becomes an  
increasing function of
$\Delta m^2$ when
the latter increases further up to
$\Delta m^2 \simeq 8 \cdot 10^{-3}~\mathrm{eV}^2$,
where $\sin^{2} \theta < 2 \cdot 10^{-2}$.
Somewhat weaker constraints on  $\sin^2\theta$
have been obtained by the Palo Verde 
collaboration \cite{PaloV}.
In the future, $\sin^2\theta$ might be 
further constrained or determined,
e.g., in long baseline neutrino 
oscillation experiments \cite{MINOS}.

\indent \hskip 1.0truecm The long baseline experiment with reactor
$\bar{\nu}_e$ KamLAND \cite{Piepke:2001tg}
has been designed 
to test the LMA MSW solution 
of the solar neutrino problem.
This experiment is planned to provide
a rather precise measurement 
of $\ms$ and $\ts$.
Due to the long baseline of the 
experiment, $L \sim 180$ km, however, 
$\ms$ can be determined with a relatively
good precision only if $\ms \ltap 10^{-4}$ eV$^2$.

\indent \hskip 1.0truecm The explanation of both the atmospheric and 
solar neutrino data in terms of 
neutrino oscillations requires,
as is well-known,
the existence of 3-neutrino mixing
in the weak charged lepton 
current: 
\begin{equation}
\nu_{l \mathrm{L}}  = \sum_{j=1}^{3} U_{l j} \nu_{j \mathrm{L}},
\label{3numix}
\end{equation}
\noindent where $\nu_{lL}$, $l  = e,\mu,\tau$,
are the three left-handed flavour 
neutrino fields,
$\nu_{j \mathrm{L}}$ is the 
left-handed field of the 
neutrino $\nu_j$ having a mass $m_j > 0$
and $U$ is a $3 \times 3$ unitary mixing matrix - 
the Pontecorvo-Maki-Nakagawa-Sakata (PMNS)
neutrino mixing matrix \cite{BPont57,MNS62}.
The three neutrino masses $m_{1,2,3}$ can obey the 
so-called normal hierarchy (NH) relation
$m_1 < m_2 < m_3$,
or that of the inverted hierarchy (IH)
type, $m_3 < m_1 < m_2$.
Thus, in order to reconstruct 
the neutrino mass spectrum 
in the case of 3-neutrino mixing,
it is necessary to 
establish, in particular,
which of the two possible 
types of neutrino mass spectrum
is actually realized.  
This information is particularly important  
for the studies of a number of fundamental
issues related to lepton mixing, 
as like the possible Majorana 
nature of massive neutrinos,
which can manifest itself
in the existence of
neutrino-less double $\beta$-decay 
(see, e.g., \cite{BilPet87,BilPetPas}).
It would also constitute a critical test 
for theoretical models of
fermionic mass matrices and 
flavor physics in general.

\indent  \hskip 1.0truecm It would be possible to 
determine whether the neutrino 
mass spectrum is with normal 
or inverted hierarchy in terrestrial 
neutrino oscillation experiments with a 
sufficiently long baseline, so that the 
neutrino oscillations take place in the Earth and 
the Earth matter effects in the oscillations
are non-negligible \cite{Lipari00,Barger00,Freund00}. 
The ambiguity regarding the type of the 
neutrino mass spectrum might be resolved
by the MINOS experiment \cite{MINOS},
although on the baseline of this experiment
the matter effects are 
relatively small \cite{Lipari00}. 
This might be done in an experiment 
with atmospheric neutrinos,
utilizing a detector 
with a sufficiently good muon charge
discrimination \cite{Pepe}.
The experiments
at neutrino factories would be
particularly suitable for the 
indicated purpose \cite{Barger00,Freund00}.

\hskip 1.0truecm    In this paper, in the context of 
three-neutrino oscillations,
we study the possibility of using anti-neutrinos
from nuclear reactors to explore 
the $\ms > 10^{-4}$ eV$^2$ region of
the LMA MSW solution.
Such an experiment 
might be of considerable interest
if, in particular,
the results of the KamLAND experiment will
confirm the validity of the 
LMA-MSW solution of the 
solar neutrino problem, but will 
allow to obtain only a lower bound 
on $\ms$ due to the fact that 
$\ms > 10^{-4}$ eV$^2$ 
\cite{BarbStru00,SchoenertTAUP01,StruViss01}.
We determine the optimal baseline of
the possible experiment with reactor $\bar{\nu}_e$,
which would provide a precise measurement of 
$\ms$ in the region  $10^{-4}~{\rm eV^2} < \ms 
\ltap 8\times 10^{-4}~{\rm eV^2}$. 
Furthermore, the same experiment might be used
to try to distinguish 
between the two types of neutrino mass 
spectrum - with normal or with inverted
hierarchy. This might be done
by exploring the effect 
of interference between the 
amplitudes of neutrino oscillations,
driven by the solar and 
atmospheric $\Delta m^2$, i.e., by 
$\ms$ and $\ma$. 
For the optimal baseline found 
earlier, $L \cong (20 - 25)$ km,
the indicated effect could be relevant
for $10^{-4}~{\rm eV^2} < \ms 
\ltap 5\times 10^{-4}~{\rm eV^2}$.
For larger values of $\ms$
within the interval (\ref{sollimit}),
the effect could be relevant
at $L \cong 10$ km.
Distinguishing 
between the two possible types of 
neutrino mass spectrum requires
a relatively high precision measurement of the
positron spectrum in the reaction
$\bar{\nu}_e + p \rightarrow e^{+} + n$
(i.e., a high statistics experiment with
sufficiently good energy resolution), 
a measurement 
of $\ma$ with very high precision, 
$\ts \neq 1.0$, e.g.,
$\ts \ltap 0.9$, and a sufficiently 
large value of the angle $\theta$,
which for $\ms \ll \ma$
controls, e.g., the 
oscillations of the atmospheric
$\nu_e$ and $\bar{\nu}_e$ and is
constrained by the CHOOZ and Palo Verde data. 

%
\section{The $\bar{\nu}_e$ Survival Probability}
\label{section:formulae}
\vspace{0.3cm}
%
\indent \hskip 1.0truecm 
We shall assume in what follows that the 3-neutrino mixing
described by eq. (\ref{3numix}) takes place. 
We shall number (without loss of generality) 
the neutrinos with definite mass in vacuum 
$\nu_j$, $j=1,2,3$, 
in such a way that their masses obey 
$m_1 < m_2 < m_3$. 
Then the cases of NH and 
IH neutrino mass spectrum
differ, in particular,
by the relation between
the mixing matrix elements 
$|U_{e j}|$, $j=1,2,3$, and 
the mixing angles 
$\theta_{\odot}$ and $\theta$
(see further).
With the indicated choice one has 
$\Delta m^2_{jk} > 0$ for $j > k$.
Let us emphasize that we do not assume
any of the relations
$m_1 \ll m_2 \ll m_3$, or 
$m_1 \ltap m_2 \ll m_3$,
or $m_1 \ll m_2 \cong m_3$, 
to be valid in what follows. 

\indent \hskip 1.0truecm Under the conditions of the experiment
we are going to discuss, which must have a baseline
$L$ considerably shorter than the baseline 
$\sim 180$ km of the KamLAND experiment,
the reactor $\bar{\nu}_e$ oscillations will 
not be affected by Earth matter effects
when the $\bar{\nu}_e$ travel 
between the source (reactor) and the detector.
If 3-neutrino mixing takes place, eq.~(\ref{3numix}), the 
$\bar{\nu}_e$ would take part in 3-neutrino
oscillations in vacuum on the way to the
detector. 

\indent \hskip 1.0truecm We shall obtain next 
the expressions for the reactor
$\bar{\nu}_e$ survival probability of interest 
in terms of measurable quantities for the two types 
of neutrino mass spectrum.
In the case of normal hierarchy between the neutrino masses
we have: 
\begin{equation}
 \Delta m^2_{\odot}= \Delta m^2_{21},~~~ 
\label{dm21sol}
\end{equation}
%
\noindent and 
\begin{equation}
|U_{\mathrm{e} 1}| = \cos \theta_{\odot} \sqrt{1 - |U_{\mathrm{e} 3}|^2},
~~|U_{\mathrm{e} 2}| = \sin \theta_{\odot} \sqrt{1 - |U_{\mathrm{e}3}|^2},
\label{Ue12}
\end{equation}
%
\noindent where 
\begin{equation}
\theta_{\odot} = \theta_{12},~~
|U_{e3}|^2 = \sin^2\theta \equiv \sin^2\theta_{13},
\label{Ue3th13}
\end{equation}
%
\noindent $\theta_{12}$ and $\theta_{13}$ being 
two of the three mixing
angles in the standard parameterization of
the PMNS matrix (see, e.g., \cite{BGG99}). 
Note that $|U_{e3}|^2$ is constrained by 
the CHOOZ and Palo Verde results.
It is not difficult to derive the expression 
for the $\bar{\nu}_e$ 
survival probability in the case under discussion:
\bea
\lefteqn{P_{NH}({\bar \nu_e}\to{\bar \nu_e})} \nonumber\\
&& =\,~~ 1 - 2 \, \sin^2\theta \cos^2\theta\,
\left( 1 - \cos \frac{ \Delta{m}^2_{31} \, L }{ 2 \, E_{\nu} } \right)
\nonumber \\
&& - \,~~\frac{1}{2} \cos^4\theta\,\sin ^{2}2\theta_{\odot} \,
\left( 1 - \cos \frac{ \Delta{m}^2_{\odot} \, L }{ 2 \, E_{\nu} } \right) 
\label{P21sol}  \\
& & +\,~~ 2\,\sin^2\theta\,\cos^2\theta\, \sin^{2}\theta_{\odot}\, 
\left(\cos
\left( \frac
{\Delta{m}^2_{31} \, L }{ 2 \, E_{\nu}} - \frac {\Delta{m}^2_{\odot} \, L }{ 2 \,
E_{\nu}}\right)
-\cos \frac {\Delta{m}^2_{31} \, L }{ 2 \, E_{\nu}} \right)\, ,
\nonumber
\eea
%
\noindent where $E_{\nu}$ is the neutrino energy and we have
made use of eqs. (\ref{dm21sol}), (\ref{Ue12}) and (\ref{Ue3th13}).

\hskip 1.0truecm  If the neutrino mass spectrum is 
with inverted hierarchy one has (see, e.g., \cite{BGKP96,BilPetPas,BGG99}):
\begin{equation}
 \Delta m^2_{\odot}= \Delta m^2_{32},~~~ 
\label{32sol}
\end{equation}
%
\noindent and 
\begin{equation}
|U_{\mathrm{e} 2}| = \cos \theta_{\odot} \sqrt{1 - |U_{\mathrm{e} 1}|^2},
~~|U_{\mathrm{e} 3}| = \sin \theta_{\odot} \sqrt{1 - |U_{\mathrm{e} 1}|^2}.
\label{Ue23}
\end{equation}
%
\noindent 
The mixing matrix element
constrained by the CHOOZ and Palo Verde data
is now $|U_{\mathrm{e} 1}|^2$ :
\begin{equation}
|U_{e1}|^2 = \sin^2\theta.
\label{Ue1th}
\end{equation}
%
The expression for the $\bar{\nu}_e$ 
survival probability can be written in the form~\cite{BNP01}:
\bea
\lefteqn{P_{IH}({\bar \nu_e}\to{\bar \nu_e})} \nonumber\\
&& =\,~~ 1 - 2 \, \sin^2\theta\,\cos^2\theta\,
\left( 1 - \cos \frac{ \Delta{m}^2_{31} \, L }{ 2 \, E_{\nu} } \right)
\nonumber \\
&& - \,~~\frac{1}{ 2} \cos^{4}\theta\,\sin ^{2}2\theta_{\odot} \,
\left( 1 - \cos \frac{ \Delta{m}^2_{\odot} \, L }{ 2 \, E_{\nu} } \right) 
\label{P32sol}
  \\
& & + \,~~2\,\sin^2\theta\,\cos^2\theta\, \cos^{2}\theta_{\odot} \,  
\left(\cos
\left( \frac
{\Delta{m}^2_{31} \, L }{ 2 \, E_{\nu}} - \frac {\Delta{m}^2_{\odot} \, L }{ 2 \,
E_{\nu}}\right)
-\cos \frac {\Delta{m}^2_{31} \, L }{ 2 \, E_{\nu}} \right)\,. \nonumber
\eea
%
 
\indent \hskip 1.0truecm Several comments concerning the expressions 
for the $\bar{\nu}_e$ survival probability,
eqs. (\ref{P21sol}) and (\ref{P32sol}),
follow.  In the first lines in the right-hand 
side of eqs. (\ref{P21sol}) and (\ref{P32sol}),
the oscillations of the 
electron (anti-)neutrino driven by
the ``atmospheric'' $\Delta m^2_{31}$
are accounted for. The CHOOZ and 
Palo Verde experiments
are primarily sensitive to this term and their  
results limit $\sin^2\theta$.
The second lines in the expressions in 
eqs. (\ref{P21sol}) and (\ref{P32sol})
contain the solar neutrino oscillation parameters.
This is the term KamLAND should be most sensitive to. 
For $\Delta m^2_{\odot} \ll 
\Delta m^2_{31} = \ma$, 
$\Delta m^2_{\odot} \ltap 10^{-4}~{\rm eV^2}$,
only one of the indicated two 
terms leads to an oscillatory dependence of
the  $\bar{\nu}_e$ survival probability
for the ranges of $L/E_{\nu}$ characterizing the
CHOOZ and Palo Verde, and the KamLAND experiments:
on the source-detector distance $L$ of the 
CHOOZ and Palo Verde experiments 
the oscillations due to $\ms$ cannot develop,
while on the distance(s) traveled by the 
$\bar{\nu}_e$ in the KamLAND experiment 
$\ma$ causes fast oscillations
which average out and are not predicted to lead,
e.g., to specific spectrum distortions
of the KamLAND event rate.

\hskip 1.0truecm The terms in the third lines  
in eqs. (\ref{P21sol}) and (\ref{P32sol})
are not present in any two-neutrino oscillation
analysis. They represent 
interference terms between the 
amplitudes of neutrino oscillations,
driven by the solar and 
atmospheric neutrino mass squared differences. 
The term in eq.~(\ref{P21sol}) is proportional to 
$\sin^2 \theta_{\odot}$, 
while the corresponding term in eq.~(\ref{P32sol})
is proportional to $\cos^2 \theta_{\odot}$ \cite{BNP01}. 
This is the only difference between
$P_{NH}({\bar \nu_e}\to{\bar \nu_e})$ 
and $P_{IH}({\bar \nu_e}\to{\bar \nu_e})$,
that can be used to distinguish between the two cases
of neutrino mass spectrum in an experiment
with reactor $\bar{\nu}_e$.
Obviously, if $\cos2\theta_{\odot} = 0$,
we have $P_{NH}({\bar \nu_e}\to{\bar \nu_e})
= P_{IH}({\bar \nu_e}\to{\bar \nu_e})$ 
and the two types of spectrum
would be indistinguishable
in the experiments under discussion. 
For vanishing $\sin^2\theta$, only the terms in the 
second line of eqs. (\ref{P21sol}) and (\ref{P32sol})
survive, and the 
two-neutrino mixing formula for solar neutrino oscillations
in vacuum is exactly reproduced.

\hskip 1.0truecm  Let us discuss next the ranges of values the 
different oscillation parameters,
which enter into the expressions
for the probabilities of interest
$P_{NH}({\bar \nu_e}\to{\bar \nu_e})$ 
and $P_{IH}({\bar \nu_e}\to{\bar \nu_e})$,
can take. The allowed region of values of $\Delta m^2_{31}$,
$\ms$, $\sin^2\theta_{\odot}$ and $\theta$
should be determined in a global 
3-neutrino oscillation analysis of the solar,
atmospheric and reactor neutrino oscillation data,
in which, in particular, $\ms$ should be allowed 
to take values in the
LMA solution region, including the interval
$\ms \sim (1.0 - 6.0)\times 10^{-4}~{\rm eV^2}$.
Such an analysis is lacking in the literature.
However, as was shown in \cite{Fogli3nusol},
a global analysis of the indicated type
would not change essentially
the results for the LMA MSW solution
we have quoted
\footnote{Let us note that the LMA MSW 
solution values of $\ms$ and $\theta_{\odot}$
we quote in eqs. (\ref{sollimit})  
and (\ref{thetalimit}) were obtained 
by taking into account
the CHOOZ and Palo Verde limits as well.}
in eqs. (\ref{sollimit})  
and (\ref{thetalimit}) as long as 
$\Delta m^2_{31} \gtap 1.5\times 10^{-3}~{\rm eV^2}$.
The reason is that for 
$\Delta m^2_{31} \gtap 1.5\times 10^{-3}~{\rm eV^2}$
and $\ms \ltap 6.0\times 10^{-4}~{\rm eV^2}$,
the solar $\nu_e$ survival probability,
which determines the level 
of suppression of the
solar neutrino flux and plays a major role
in the analyses of the solar neutrino data,
depends very weakly on
(i.e., is practically independent of) 
$\Delta m^2_{31}$. Thus, $\ms$ and $\theta_{\odot}$
are uniquely determined by
the solar neutrino and CHOOZ and Palo Verde 
data, independently of the 
atmospheric neutrino data
and of the type of the neutrino mass spectrum.
The CHOOZ and Palo Verde data lead to an
upper limit on $\ms$ in the LMA MSW solution region
(see, e.g., \cite{ConchaSNO,Gonza3nu}):
$\ms \ltap 7.5\times 10^{-4}~{\rm eV^2}$.
For $\ms \ltap 1.0\times 10^{-4}~{\rm eV^2}$, 
the CHOOZ and solar neutrino data 
imply the upper limit on $\sin^2\theta$ given in
eq. (\ref{chooz1}). For $\ms \sim (2.0 - 6.0)\times 10^{-4}~{\rm eV^2}$
of interest, the upper limit on $\sin^2\theta$
as a function of $\Delta m^2_{31} \gtap 10^{-3}~{\rm eV^2}$ 
for given $\ms$ and $\sin^22\theta_{\odot}$
is somewhat more stringent \cite{BNP01}.

\hskip 1.0truecm  Would a global 3-neutrino oscillation 
analysis of the solar,
atmospheric and reactor neutrino oscillation data
lead to drastically different results for
$\Delta m^2_{31}$ in the two cases of 
normal and inverted neutrino mass 
hierarchy? Our preliminary 
analysis shows that given the 
existing atmospheric neutrino data from 
the Super-Kamiokande experiment,
such an analysis  
i) would not be able
to discriminate  between the two cases
of neutrino mass spectrum, and
ii) would give essentially 
the same allowed region for $\Delta m^2_{31}$
in the two cases of neutrino mass spectrum.
We expect the regions of allowed values of 
the mixing angle $\theta_{atm}$, which 
controls the dominant atmospheric  
$\nu_{\mu}\rightarrow \nu_{\tau}$
and $\bar{\nu}_{\mu}\rightarrow \bar{\nu}_{\tau}$
oscillations, to differ somewhat in the two cases.
Note, however, that this mixing angle 
does not enter the expression for the 
$\bar{\nu}_e$ survival probability
we are interested in.

\hskip 1.0truecm For $\ms \ltap 1.0\times 10^{-4}~{\rm eV^2}$
and sufficiently small values of $\sin^2\theta$,
$\Delta m^2_{31}$ coincides effectively
with $\Delta m^2_{atm}$ of the two-neutrino 
$\nu_{\mu}$ and $\bar{\nu}_{\mu}$ 
oscillation analyses of the SK 
atmospheric neutrino data.
If $\sin^2\theta > 0.01$, a three-neutrino 
oscillation analysis of the atmospheric neutrino 
and CHOOZ data, performed under the assumption of
$\ms \ltap 1.0\times 10^{-4}~{\rm eV^2}$ \cite{Gonza3nu}, 
gives regions of allowed values of  
$\Delta m^2_{atm} = \Delta m^2_{31}$,
which are correlated with 
the value of $\sin^2\theta$. The latter
must satisfy the CHOOZ and Palo Verde 
constraints. 

\hskip 1.0truecm At present, as we have already indicated,
a complete three-neutrino oscillation analysis
of the atmospheric neutrino and CHOOZ
data with $\ms$ allowed to take values 
up to $\sim (6.0 - 7.0)\times 10^{-4}~{\rm eV^2}$,
i.e., in the region where deviations from the two-neutrino
approximation could be non-negligible, 
is lacking in the literature.
Therefore in what follows we will use 
representative values of
$\Delta m^2_{31}$ which lie in the region
given by eq. (\ref{atmlimit}).

\section{The Difference between $P_{NH}({\bar \nu_e}\to{\bar \nu_e})$ 
and $P_{IH}({\bar \nu_e}\to{\bar \nu_e})$}  
\vspace{0.3cm}
%

\indent \hskip 1.0truecm Let us discuss next in greater detail the difference
between the $\bar{\nu}_e$ surviving probabilities
in the two cases of neutrino mass spectrum of interest,
$P_{NH}({\bar \nu_e}\to{\bar \nu_e})$ 
and $P_{IH}({\bar \nu_e}\to{\bar \nu_e})$.
While the terms in the first two lines
in eqs. (\ref{P21sol}) and (\ref{P32sol})
describe oscillations in $L/E_{\nu}$ with frequencies
$\Delta m^2_{31}/4 \pi$ and $\ms/4 \pi$, respectively, 
the third term has the shape of
beats, being produced by the 
interference of two waves, with
the same amplitude but slightly different 
frequencies:
\bea
\cos
\left( \frac
{\Delta{m}^2_{31} \, L }{ 2 \, E_{\nu}} - \frac {\Delta{m}^2_{\odot} \, L }{ 2 \,E_{\nu}}
\right)
-\cos \frac {\Delta{m}^2_{31} \, L }{ 2 \, E_{\nu}}\,
&=&
2\,\sin  \, \frac {\Delta{m}^2_{\odot} \, L }{ 4 \,E_{\nu}}
\, \sin \,  \left( \frac
{\Delta{m}^2_{31} \, L }{ 2 \, E_{\nu}} - \frac {\Delta{m}^2_{\odot} \, L }{ 4 \,E_{\nu}}
\right)
\nonumber \\
&\simeq& 2\,\sin  \, \frac {\Delta{m}^2_{\odot} \, L }{ 4 \,E_{\nu}}
\, \sin \,  \left( \frac
{\Delta{m}^2_{31} \, L }{ 2 \, E_{\nu}}
\right)
\eea
This is a modulated oscillation
with approximately the same 
frequency of the first term in 
eqs. (\ref{P21sol}) and (\ref{P32sol})
($\Delta m^2_{31}/4 \pi$)
and amplitude oscillating
between $0$ and $2\sin^2 \theta_{\odot}$ of the amplitude 
of the first term  itself. The  beat frequency is equal to the 
frequency of the dominant oscillation  ($\ms/4 \pi$).   
The modulation is exactly in phase with the $\ms-$driven 
dominant oscillation of interest, so that the maximum of the 
oscillation amplitude of the 
interference term (third lines in the expressions for 
$P_{NH}({\bar \nu_e}\to{\bar \nu_e})$ 
and $P_{IH}({\bar \nu_e}\to{\bar \nu_e})$)
is reached in coincidence with the points of 
maximal decreasing of the $\bar{\nu}_e$ survival probability,
where $\ms\,L/4\,E = \pi/2$, and vice versa -
this amplitude vanishes at the local maxima
of the survival probability.
At the minima of the $\bar{\nu}_e$ 
survival probability, for instance at $\ms\,L/4\,E_{\nu} = \pi/2$,
$P^{NH(IH)}\,({\bar \nu_e}\to{\bar \nu_e})$
takes the value:

\bea
P^{NH(IH)}\,({\bar \nu_e}\to{\bar \nu_e})
\Big|_{\frac{\ms L}{2\pi E_{\nu}}=1}& =&
1-2\,\sin^2\theta\,\cos^2\theta-\cos^4\theta\,\sin^22\theta_{\odot}
\nonumber \\
&& 
^{(+)}_{~-}\,\cos2\theta_{\odot}\,
  2\, \sin^2\theta\,\cos^2\theta\, \cos\,\pi\frac{\Delta m^2_{31}}{\ms} 
\,. 
\label{pi/2}
\eea

From eqs. (\ref{P21sol}), (\ref{P32sol}) and (\ref{pi/2})
one deduces that:
\begin{itemize}

\item for maximal mixing, $\cos2\theta_{\odot} = 0$,
the last term cancels, and 
$P^{NH}=P^{IH}$;

\item for very small mixing angles, $\cos2\theta_{\odot}\simeq 1$,
the terms describing the oscillations 
driven by $\Delta m^2_{31}$ 
in the NH and IH cases 
have opposite signs: the two waves are 
exactly out of phase.

\item for intermediate values of $\cos2\theta_{\odot}$ from the LMA
MSW solution region, $\cos2\theta_{\odot} \cong (0.3 - 0.6)$,
the $\Delta m^2_{31}-$driven 
contributions in the cases of normal and inverted hierarchy 
have still opposite signs and the magnitude of the effect
is proportional to $2\cos2\theta_{\odot} \sin^2\theta$.

\end{itemize}

\hskip 1.0truecm The net result of these 
properties is that 
in the region of the minima of the 
$\bar{\nu}_e$ survival probability 
due to $\ms$, where $\ms\,L\,/(2E) = \pi (2k + 1)$,
$k = 0,1,\dots$,
the difference between 
$P_{NH}({\bar \nu_e}\to{\bar \nu_e})$ 
and $P_{IH}({\bar \nu_e}\to{\bar \nu_e})$)
is maximal.
In contrast, at the maxima of 
$P_{NH}({\bar \nu_e}\to{\bar \nu_e})$ 
and $P_{IH}({\bar \nu_e}\to{\bar \nu_e})$)
determined by $\ms\,L\,/(2E) = 2\pi k$,
we have, for any $\sin^2\theta_{\odot}$,
$P_{NH}({\bar \nu_e}\to{\bar \nu_e}) =
P_{IH}({\bar \nu_e}\to{\bar \nu_e})$.

\hskip 1.0truecm The two-neutrino oscillation approximation
used in the analysis of 
the CHOOZ and Palo Verde data
is rather accurate 
as long as $\ms$ is sufficiently small~\cite{BNP01}: 
for $\ms \ltap 10^{-4}~{\rm eV^2}$,
the $L/E_{\nu}$ values characterizing 
these experiments, chosen to ensure
maximal sensitivity to $\Delta m^2_{31} \gtap 10^{-3}~{\rm eV^2}$,
are much smaller than the value  
at which the first minimum of 
$P_{NH(IH)}({\bar \nu_e}\to{\bar \nu_e})$
due to the $\ms$-dependent 
oscillating term occurs. 
Correspondingly, the effect of the interference
term is strongly suppressed by the beats.
For $\ms \gtap 2\times 10^{-4}~{\rm eV^2}$ 
this is no longer valid and the 
interference term under discussion 
has to be taken into account
in the analyses of the CHOOZ and Palo Verde data
\cite{BNP01}.
  
%
\section{Measuring Large $\ms$ at Reactor Facilities}
\label{section:ms}
\vspace{0.2cm}
%

\hskip 1.0truecm As is well-known,
nuclear reactors are intense sources of low 
energy $\bar{\nu}_e$ ($E_{\nu} \ltap 8$ MeV), emitted 
isotropically in the $\beta$-decays of 
fission products with high neutron density
\cite{BP46}.  Anti-neutrinos can
then be detected through the positrons produced by 
inverse $\beta$-decay on nucleons. 
The reactor $\bar{\nu}_e$ energy spectrum 
has been accurately 
measured and is theoretically well understood
~\footnote{By reactor $\bar{\nu}_e$ energy spectrum
we mean here and in what follows
the product of the $\bar{\nu}_e$ production spectrum 
and the inverse $\beta$-decay cross-section,
which gives the ``detected'' neutrino 
spectrum in the no oscillation case.
The $\bar{\nu}_e$ production 
spectrum is known with 
larger uncertainties
at $\bar{\nu_e}$ energies 
$E_{\nu} \ltap 2$ MeV, but this range is 
not of interest due to the 
threshold energy  $E_{\nu}^{th} \cong 1.8$ MeV of the 
inverse $\beta$-decay reaction~\cite{Kopeikin:2001rj}. 
Certain known time dependence 
at the level of a few percent is also
present up to 3.5 MeV~\cite{Kopeikin:2001qv} 
and should possibly be taken into account
in the analysis of the experimental data.
}
~\cite{reactors}: it
essentially consists of a bell-shaped 
distribution in energy
centered around $E_{\nu} \sim 4$ MeV, 
having a width of approximately $3$ MeV.
CHOOZ, Palo Verde and
KamLAND are examples of experiments 
with reactor $\bar{\nu}_e$,
the main difference being the distance 
between the source and the detector 
explored ($L \sim 1$ km for CHOOZ and Palo Verde,
and $L \sim 180$ km for KamLAND).

\hskip 1.0truecm The best sensitivity to a 
given value of $\Delta m^2_{\odot}$ of 
the experiment of interest is at $L$ 
at which the maximum reduction
of the survival probability is realized. 
As can be seen from eqs.~(\ref{P21sol}) - (\ref{P32sol}), this happens
for $L$ around $L^{\ast} \equiv 2\pi\,E_{\nu}/ \Delta m^2_{\odot}$. 
This implies that for 
$E_{\nu} = 4$ MeV, the optimal length to 
test neutrino oscillations
with reactor experiments is:
\bea
L^{\ast} \cong \frac {5 \times 10^{-3}} 
{(\Delta m^2_{\odot} / \mbox{eV}^2)}~\mbox{km}
\eea
The best sensitivity of KamLAND, for instance, is in the 
range of $2 \div 3 \times 10^{-5}$ eV$^2$.
We will discuss next in 
greater detail the distances $L$
which could be used to probe the LMA MSW solution
region at $\ms > 10^{-4}$ eV$^2$,
in
order to extract  $\ms$ from these oscillation experiments.
%
\subsection{Total Event Rate Analysis}
\label{subsection:tot}
\begin{figure}
\begin{center}
\epsfxsize = 10cm
\epsffile{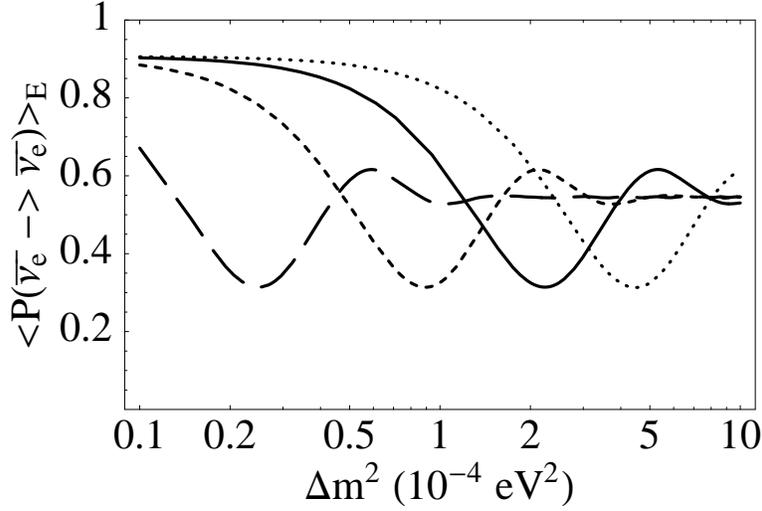}
\leavevmode
\end{center}
\caption{The reactor $\bar{\nu}_e$ survival probability,
averaged over the $\bar{\nu}_e$ energy spectrum, for 
$\Delta m^2_{31} = 2.5 \times 10^{-3}$ eV$^2$, 
$\ts = 0.8$, $\sin^2\theta=0.05$, as
a function of $\ms$. The curves correspond to $L=180$ km (long dashed), 
$L=50$ km (dashed),  $L=20$ km (thick) and $L=10$ km (dotted), respectively.
}
\label{totalrate}
\end{figure}

%
\hskip 1.0truecm One of the signatures of the $\bar{\nu}_e-$oscillations
would be a substantial reduction
of the measured total event rate due to the
reactor $\bar{\nu}_e$ in comparison
with the predicted one in the absence 
of oscillations. 
In order to compute 
the expected total event rate one has to 
integrate  the $\bar{\nu}_e$ survival probability 
multiplied by the $\bar{\nu}_e$ energy spectrum 
over $E_{\nu}$. In Fig. \ref{totalrate} we show this 
averaged survival probability for 
different values of $L$ as a function of $\ms$, using 
the ``best fit'' values \cite{SKatm,FogliSNO,ConchaSNO,GoswaSNO}
for $\Delta m^2_{31}$ and $\ts$.
%
%
%
When averaging over the $\bar{\nu}_e$ energy spectrum, 
oscillatory effects with too short a period are washed out,
and the experiment is sensitive only to the average amplitude.
This happens when the width $\delta E_{\nu}$ of the energy spectrum 
is such that the integration runs over more than one period, i.e.,
approximately for:
\bea
\label{width}
\delta E_{\nu} \gtap \frac{4\pi\,E_{\nu}^2}{\Delta m^2\,L} \simeq 
\frac{4 \times 10^4 \,\mbox{eV}^3}{\Delta m^2\, {\mbox(L/Km)}}.
\eea
%
Since $\delta E_{\nu} \sim 3$ MeV, at KamLAND this happens 
approximately for 
$\Delta m^2_{\odot} \gtap 7 \times 10^{-5}$ eV$^2$. The corresponding 
curve in Fig. 1 indicates 
that the actual sensitivity 
extends to somewhat larger
values of $\ms$ than what is expected
on the basis on the above estimate,
but the total event rate becomes flat for  
$\Delta m^2_{\odot} \gtap  10^{-4}$ eV$^2$. This means
that KamLAND will be able, through the 
measurement of the total even rate,
to test all the region of the LMA MSW solution 
and determine whether the latter
is the correct 
solution of the solar neutrino problem, 
but will provide a precise measurement 
of $\ms$ only if 
$\ms \ltap 10^{-4}$ eV$^2$. 
If $\ms \gtap 2\times 10^{-4}$ eV$^2$,
it would be possible to obtain
only a lower bound on $\ms$
and a new experiment might be required
to determine $\ms$.

   Fig. 1 shows that as $L$ decreases, 
the sensitivity region moves to larger $\ms$.
These results imply that 
a reactor $\bar{\nu}_e$  
experiment with $L \cong (20 - 25)$ km 
can probe the  range 
$0.8\times 10^{-4}\, \mbox{eV}^2 < \ms  
\ltap 6 \times  10^{-4}\, \mbox{eV}^2$.
One finds that for $\ms \cong 2 \times 10^{-4}$ eV$^2$ and
$\Delta m^2_{31} \cong  2.5\times 10^{-3}$ eV$^2$,
the best sensitivity is at $L \cong 20$ km. 
Moreover, with $L \cong (20 - 25)$ km, 
the predicted total event rate deviates from being 
flat (in $\ms$) 
actually for $\ms$ as large as
$\sim (5 - 6)\times  10^{-4}$ eV$^2$.
In order to have a precise determination of
$\ms$ with $L \cong (20 - 25)$ km
for the largest 
values given in eq.~(\ref{sollimit}),
$\ms \cong (7 \div 8) \times 10^{-4}$
eV$^2$, one should use the information about
the $e^{+}-$spectrum distortion
due to the $\bar{\nu}_e-$oscillations.
By measuring the  $e^{+}-$spectrum
with a sufficient precision
it would be possible to cover the whole interval
\bea
1.0\times 10^{-4}\, \mbox{eV}^2 \ltap \ms  \ltap 
8.0 \times  10^{-4}\, \mbox{eV}^2~,
\eea 
%
i.e., to determine $\ms$ if it lies in this interval,
by performing an experiment at
$L \cong (20 - 25)$ km from the reactor(s)
\footnote{The fact that if
$\ms \cong 3.2 \times 10^{-4}$ eV$^2$,
a reactor $\bar{\nu}_e$  
experiment with $L \cong 20$ km 
would allow to measure $\ms$ 
with a high precision was also noticed recently 
in~\cite{StruViss01}.}
(see the next sub-section).

\hskip 1.0truecm Applying eq. (17) with 
$\Delta m^2 = \Delta m^2_{31}$, 
one sees that for the ranges of $L$
which allow to probe $\Delta m^2_{\odot}$ 
from the LMA MSW solution region,
the total event rate is not sensitive to the
oscillations driven by 
$\Delta m^2_{31} \gtap 1.5\times 10^{-3}~{\rm eV^2}$.
Thus, the total event rate analysis  
would determine $\ms$ which would be the same for
both the normal and inverted hierarchy 
neutrino mass spectrum.

%
\subsection{Energy Spectrum Distortions}
\label{subsection:energy}
%
\hskip 1.0truecm An unambiguous evidence 
of neutrino oscillations would
be the characteristic distortion of the
$\bar{\nu}_e$ energy spectrum. This is caused
by the fact that, at fixed $L$, neutrinos with 
different energies
reach the detector in a different 
oscillation phase, so that
some parts of the spectrum would be suppressed more strongly
by the oscillations than other parts.
The search for distortions 
of the $\bar{\nu}_e$ energy spectrum 
is essentially a direct test of the 
$\bar{\nu}_e$ oscillations. It is more effective
than the total rate analysis since it is not 
affected, e.g., by the overall normalization of the
reactor $\bar{\nu}_e$ flux. 
However, such a test requires a sufficiently high 
statistics and sufficiently good energy
resolution of the detector used. 

\begin{figure}
\begin{center}
\epsfxsize = 12cm
\epsffile{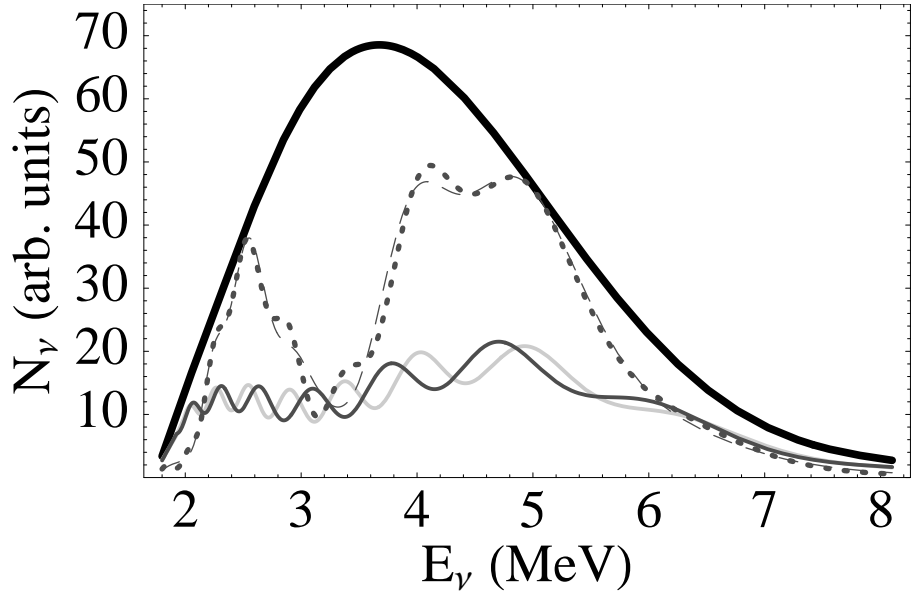}
\leavevmode
\end{center}
\caption{The reactor $\bar{\nu}_e$ energy spectrum
at distance $L=20$ km from the source, 
in the absence of $\bar{\nu}_e$ oscillations 
(double-thick solid line) and in the case of 
$\bar{\nu}_e$ oscillations characterized by
$\Delta m^2_{31}=2.5 \times 10^{-3}$ eV$^2$, $\ts=0.8$
and $\sin^2\theta=0.05$. 
The thick lines are obtained 
for $\ms=2 \times 10^{-4}$ eV$^2$
and correspond to NH (light grey) 
and IH (dark grey) neutrino mass spectrum. 
Shown is also the spectrum for $\ms=6 \times 10^{-4}$ eV$^2$ 
in the NH (dotted) and IH (dashed) cases.
}
\label{spectrum}
\end{figure}

\hskip 1.0truecm Energy spectrum distortions can be studied, 
in principle, in an experiment with $L \cong (20 - 25)$ km.
In Fig.~\ref{spectrum} we show the comparison between 
the $\bar{\nu}_e$ spectrum
expected for $\ms=2 \times 10^{-4}$ eV$^2$ and
$\ms=6 \times 10^{-4}$ eV$^2$ and the spectrum in the 
absence of $\bar{\nu}_e$ oscillations. 
No averaging has been performed and the
possible detector resolution 
is not taken into account. 
The curves show the product 
of the probabilities given by 
eqs.~(\ref{P21sol}) and (\ref{P32sol}) and
the predicted reactor 
$\bar{\nu}_e$ spectrum~\cite{spec}.
%
%
\hskip 1.0truecm As Fig. \ref {spectrum} illustrates,
the $\bar{\nu}_e$ spectrum 
in the case of oscillation is well 
distinguishable from that in the absence of 
oscillations. Moreover, 
for $\ms$ lying in the interval
$10^{-4}~{\rm eV^2} < \ms \ltap 8.0 \times 10^{-4}$ eV$^2$, 
the shape of the spectrum
exhibits a very strong dependence on the value
of $\ms$. A likelihood analysis 
of the data would be able to
determine the value of $\ms$ 
from  the indicated interval
with a rather good precision.
This would require a precision 
in the measurement
of the $e^{+}-$spectrum, 
which should be just not worse than the
precision achieved in the CHOOZ experiment
and that planned to be reached in the
KamLAND experiment. 
If the energy bins used in the measurement
of the spectrum are sufficiently large,
the value of $\ms$ thus determined should
coincide with value obtained from 
the analysis of the total event rate and should 
be independent of $\Delta m^2_{31}$.

%
\section{Normal vs. Inverted Hierarchy}
\label{section:vs}
\vspace{0.2cm}
%

\hskip 1.0truecm  
In Fig.~\ref{spectrum} we show 
the deformation of the reactor $\bar{\nu}_e$
spectrum both for the 
normal and inverted hierarchy 
neutrino mass spectrum: as
long as no integration over 
the energy is performed, the deformations 
in the two cases of neutrino mass spectrum 
can be considerable, and the sub-leading 
oscillatory effects driven 
by the atmospheric mass squared 
difference (see the first and the
third line of eqs.~(\ref{P21sol}) - (\ref{P32sol})) 
can, in principle, be
observed. They could 
be used to distinguish between 
the two hierarchical patterns, 
provided the solar mixing is not maximal
\footnote{It would be impossible
to distinguish between the normal and inverted
hierarchy neutrino mass spectrum
if for given $\Delta m^2_{\odot} > 10^{-4}~{\rm eV^2}$ 
and $\sin^22\theta_{\odot}\neq 1$,
the LMA solution region 
is symmetric with respect to the change
$\theta_{\odot} \rightarrow \pi/2 - \theta_{\odot}$ 
($\cos2\theta_{\odot} \rightarrow - \cos2\theta_{\odot}$).
While the value of $\sin^22\theta_{\odot}$ is expected
to be measured with a relatively high precision
by the KamLAND experiment, the sign of 
$\cos2\theta_{\odot}$ will not 
be fixed by this experiment. 
However, the $\theta_{\odot} - (\pi/2 - \theta_{\odot})$
ambiguity can be resolved by the
solar neutrino data.
Note also that the current solar neutrino data
disfavor values of $\cos2\theta_{\odot} < 0$
in the LMA solution region 
(see, e.g., \cite{FogliSNO,ConchaSNO,KSSNO}).}, 
$\sin^2\theta$ is not too small and
$\Delta m^2_{31}$ is known with high precision.
It should be clear that the possibility we will 
be discussing next
poses remarkable challenges.

\hskip 1.0truecm
The experiment under discussion
could be in principle an alternative to the 
measurement of the sign of $\Delta m^2_{31}$
in long (very long) baseline neutrino oscillation
experiments \cite{Lipari00,Barger00,Freund00} or 
in the experiments
with atmospheric neutrinos (see, e.g., \cite{Pepe}).

\hskip 1.0truecm  The  magnitude of the effect of interest 
depends, in particular,
on three factors, as we have already pointed out:

\begin{itemize}
\item the value of the solar mixing angle $\theta_{\odot}$: 
the different behavior of the two survival 
probabilities is due to the 
difference between $\sin^2 \theta_{\odot}$
and $\cos^2 \theta_{\odot}$; correspondingly, 
the effect vanishes for maximal mixing;
thus, the more the mixing deviates from the
maximal the larger the effect;

\item the value of $\sin^2\theta$, which controls 
the magnitude of the sub-leading
effects due to $\Delta m^2_{31}$ on the 
$\ms-$driven oscillations: the effect 
of interest 
vanishes in the decoupling limit 
of $\sin^2\theta \rightarrow 0$;

\item the value of $\ms$ (see Fig. \ref{totalrate}): 
for given $L$ and $\ms$ the difference 
between the spectrum in the cases of normal
and inverted hierarchy is maximal at the 
minima of the survival probability, 
and vanishes at the maxima. 
\end{itemize}

\hskip 1.0truecm
A rough estimate of the possible difference 
between the predictions of the event rate spectrum
for the two hierarchical patterns, 
is provided by the ratio between 
the difference and the sum 
of the two corresponding 
probabilities at  $\ms L=2\pi E_{\nu}$:
\bea
\frac{P_{NH}-P_{IH}}{P_{NH}+P_{IH}} =
\frac{2\,\cos 2\theta_{\odot}\,\sin^2\theta\,\cos^2\theta}
{1-2\sin^2\theta\,\cos^2\theta-\cos^4\theta\,\sin^22\theta_{\odot}}
\, \cos\,\pi\,\frac{\Delta m^2_{31}}{\ms}\,. 
\label{estimate}
\eea
%
The ratio could be rather large: the factor in front of 
the $\cos\,\pi\,\Delta m^2_{31}/\ms$
is about $25\%$ for $\ts=0.8$ and $\sin^2\theta = 0.05$.

\hskip 1.0truecm The actual feasibility 
of the study under discussion depends 
crucially on the  
integration over (i.e., the binning in) 
the energy: for the effect not to be 
strongly suppressed, the energy
resolution of the detector $\Delta E_{\nu}$ 
must satisfy:
\bea
\label{interval}
\Delta E_{\nu} \ltap \frac{4\pi\,E_{\nu}^2}{\Delta m^2_{31}\,L} \simeq 
\frac{2 \div 6 \times 10^4 \,\mbox{eV}^3}{\Delta m_{31}^2\, (L/km)}.
\eea
%
 For $L \sim 1$ km this condition 
could be satisfied for 
$\delta E_{\nu} \simeq \Delta E_{\nu}$,
but at $L \cong (15 - 20)$ Km,  
for $\Delta m^2_{31} =2.5 \times 10^{-3}$ eV$^2$  
and  $E_{\nu}$ in the interval $(3 - 5)$ MeV, one should
have $\Delta E_{\nu} \ltap 0.5$ MeV.

\hskip 1.0truecm Our discussion
so far was performed for simplicity 
in terms of the reactor $\bar{\nu}_e$
energy spectrum,
while in the experiments of interest
one measures the energy of the positron emitted 
in the inverse $\beta$-decay, $E_e$. 
The relation between 
$E_e$ and $E_{\nu}$ is well known 
(see for instance~\cite{spec}),
and, up to corrections of 
at most  few per cent, 
consists just in
a shift due to the threshold energy of the process: 
%
\begin{figure}[h,t]
\begin{center}
\epsffile{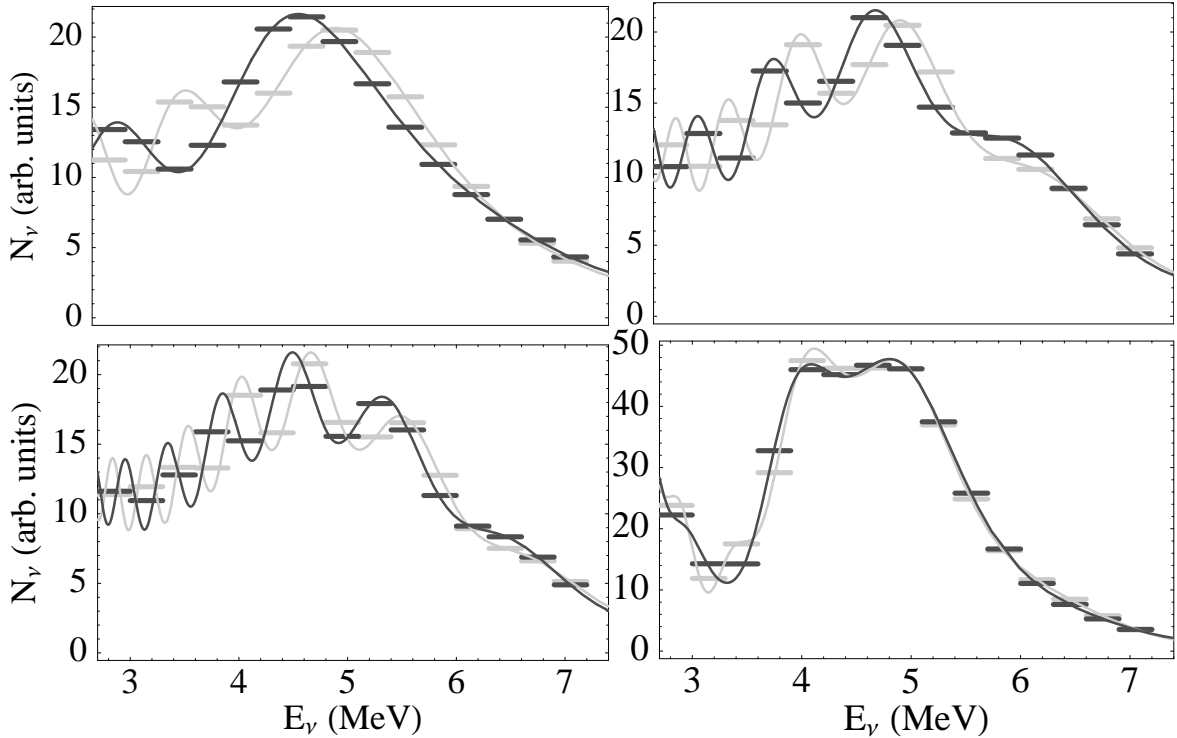}
\leavevmode
\end{center}
\caption{Comparison between the predicted 
event rate spectrum at $L=20$ km, measured in
energy bins having a width of $\Delta E_{\nu} = 0.3$ MeV
in the cases of normal (light grey) and inverted (dark grey) 
neutrino mass hierarchy. 
The two upper and the lower left figures are for $\ms = 2 
\times 10^{-4}$ eV$^{2}$,
 $\ts = 0.8$, $\sin^2\theta=0.05$, and  $\Delta
m^2_{31} = 1.3;~ 2.5;~ 3.5 \times 10^{-3}$ eV$^2$, respectively.
The lower right figure was obtained for 
$\ms = 6 \times 10^{-4}$ eV$^{2}$ and 
$\Delta m^2_{31} = 2.5 \times 10^{-3}$ eV$^2$.
}
\label{bins}
\end{figure}
%
\noindent $E_{\nu} \cong E_{e} + (E_{\nu}^{th} - m_e)$.    
The maximal $\Delta E_{\nu}$ allowed in order to make 
the effect observable can be then directly compared to 
the experimental positron energy resolution $\Delta E_e$~
\footnote{In the CHOOZ experiment, for instance,
the binning in $E_{e}$ was
$\Delta E_e \simeq 0.40$ MeV~\cite{CHOOZ}. KamLAND
is expected to have a resolution 
better than $\Delta E_e/E_e=
10\%/\sqrt{E_e}$, where $E_e$ is in 
MeV~\cite{Gonzalez-Garcia:2001zy}
.}.

\hskip 1.0truecm For $\ms \ltap 10^{-4}$ eV$^2$, 
the first (most significant) minimum of the 
survival probability can be explored if 
$L\sim 180$ km. In this 
case, due to the bigger distance $L$, 
the energy resolution required would be by a factor
of ten smaller. 
 This means that for $\ms \ll \Delta m^2_{31}$,
it is practically impossible to 
realize the condition of 
maximization of the difference
between the survival probabilities in the two 
cases of neutrino mass spectrum
without strongly suppressing the magnitude 
of the difference by 
the binning of
the energy spectrum. 

\hskip 1.0truecm  In order to illustrate what are
the concrete possibilities in the case 
of the experiment under discussion, 
we have divided the energy interval 
$2.7$ MeV $< E_{\nu} < 7.2$ MeV
into 15 bins, with $\Delta E_{\nu} = 0.3$ MeV, 
and calculated the 
value of the product
of the survival probability and the energy spectrum
in each of the bins.
The results are shown in Fig.~\ref{bins}.

\hskip 1.0truecm  As our results show and Fig. 3 indicates, 
for $\Delta m^2_{31} \cong (1.5 - 3.0) \times 10^{-3}$ eV$^2$,
%
\begin{equation}
\ms \cong (2.0 - 5.0)\times 10^{-4}~{\rm eV^2},
\label{dmsunNvsIH}
\end{equation}
%
$\ts \cong 0.8$ and $\sin^2\theta \cong (0.02 - 0.05)$,
it might be possible to distinguish 
the two cases of neutrino mass spectrum
by a high precision measurement
of the positron energy spectrum
in an experiment with reactor $\bar{\nu}_e$
with a baseline of $L \cong (20 - 25)$ km.
This should be a high
statistics experiment (not less than about 
2000 $\bar{\nu}_e-$induced events per year)
with a sufficiently good energy resolution
\footnote{Preliminary estimates show
that a detector of the type of KamLAND
and a system of nuclear reactors
with a total power of approximately
5 - 6 GW might produce the required 
statistics and precision 
in the measurement of the positron spectrum.}.
For larger values of $\ms$ not exceeding
$8.0\times 10^{-4}$ eV$^2$,
and $\Delta m^2_{31} \cong (1.5 - 3.0) \times 10^{-3}$ eV$^2$,
the experiment should be done with a smaller baseline,
$L \cong 10$ km.
If, however, $\sin^2\theta \ltap 0.01$, and/or 
$\ts \gtap 0.9$, and/or $\ts \ltap 0.9$ but 
the LMA solution admits equally 
positive and negative values of
$\cos2\theta_{\odot}$,
the difference between the spectra in the two cases
becomes hardly observable.
Further, in obtaining Fig. 3 we 
have implicitly assumed that 
$\Delta m^2_{31}$ is known with
negligible uncertainty. 
Actually, for the difference between the spectra
under discussion to be observable, 
$\Delta m^2_{31}$ has to be determined, 
according to our estimates,
with a precision of $\sim 10\%$ or better 
\footnote{The analysis, e.g, of the  MINOS potential
for a high precision determination of $\Delta m^2_{31}$
in the case of $\Delta m^2_{\odot} \ltap 10^{-4}~{\rm eV^2}$ 
yields very encouraging results (see, e.g., \cite{VBargerMINOS01}).  
For $\Delta m^2_{\odot} 
\cong (2.0 - 8.0)\times 10^{-4}~{\rm eV^2}$, 
such analysis is lacking in the literature.}: 
given the values of 
$\ms$, $\ts$ and $\sin^2\theta$,
a spectrum in the NH case
corresponding to a given $\Delta m^2_{31}$
can be rather close in shape to the spectrum
in the IH case for a different value of
$\Delta m^2_{31}$. 
There is no similar effect when varying $\ms$.

%
\section{Conclusions}
\label{section:conclusions}
%

\hskip 1.0truecm Reactor experiments 
have the possibility to 
test the LMA MSW solution of
the solar neutrino problem. While the 
KamLAND experiment 
should be able to test 
this solution, a new experiment with 
a shorter baseline might be 
required to determine 
$\ms$ with high precision if
the results of the  KamLAND experiment 
show that $\ms > 10^{-4}$ eV$^2$.
Performing a three-neutrino oscillation analysis
of both the total event rate suppression and 
the $e^{+}-$energy spectrum distortion
caused by the $\bar{\nu}_e-$oscillations in vacuum,
we show that a value of $\ms$ from the interval
$10^{-4}~{\rm eV^2} < \ms \ltap 8.0\times 10^{-4}$ eV$^2$
could be determined with a high precision 
in experiments with $L \cong (20 - 25)$ km
if the $e^{+}-$energy spectrum
is measured with a sufficiently good accuracy.
Furthermore, if $\ms \cong (1.0 - 5.0) \times 10^{-4}$ eV$^2$,
such an experiment with $L \cong (20 - 25)$ km
might also be able to distinguish between the 
cases of neutrino mass 
spectrum with normal and inverted hierarchy;
for larger values of $\ms$ not exceeding
$8.0\times 10^{-4}$ eV$^2$,
a shorter baseline, $L \cong 10$ km, should be used
for the purpose.
The indicated possibility poses remarkable challenges
and might be realized for a limited range of values of
the relevant parameters.
The corresponding detector
must have a good energy resolution 
(allowing a binning in the positron energy
with $\Delta E_{e} \ltap 0.40$ MeV) and
the observed event rate due to the reactor
$\bar{\nu}_e$ must be sufficiently high
to permit a high precision 
measurement of the $e^{+}-$spectrum.
Further, the mixing angle constrained 
by the CHOOZ and Palo Verde
data $\theta$ must be sufficiently large
($\sin^2 {\theta} \sim 0.03 - 0.05 $),
and the ``solar'' mixing angle
$\theta_{\odot}$ should not be maximal 
($\sin^2 2\theta_{\odot} \ltap 0.9$).
In addition, the value of 
$\Delta m^2_{31}$, which is 
responsible for the dominant
$\nu_{\mu} \rightarrow \nu_{\tau}$ and 
$\bar{\nu}_{\mu} \rightarrow \bar{\nu}_{\tau}$
oscillations of the atmospheric neutrinos,
should be known with a high precision. 
However, as it is well known, 
``only those who wager can win'' \cite{Pauli30}.

%
\vskip 2mm
\begin{em}
\underline {Acknowledgements}
\vspace{0.1cm}
S.T.P. would like to thank S.M. Bilenkly and D. Nicolo
for very useful discussions, and S. Schoenert for
discussions and for bringing to his attention the
reference \cite{SchoenertTAUP01}. 
S.T.P. acknowledges with gratefulness
the hospitality and support of the SLAC Theoretical 
Physics Group, where part of the work on 
the present study was done. 
This work was supported in part by the 
EEC grant ERBFMRXCT960090 and by the  
RTN European Program HPRN-CT-2000-00148.  
\end{em}
%
%
%
\begin{em}

\end{em}
\end{document}